\documentclass[runningheads]{llncs}
\usepackage[T1]{fontenc}
\usepackage{graphicx}
\usepackage{amsmath}
\usepackage{tikz}\usetikzlibrary{shapes.geometric, arrows}

\tikzstyle{process} = [rectangle, minimum width=6cm, minimum height=1cm, text centered, draw=black] % fill=blue!30] rounded corners,
\tikzstyle{arrow} = [thick,->,>=stealth]
\usepackage{comment}
\usepackage{listings}
\usepackage{xcolor}

\usepackage[english]{babel}
\usepackage[autostyle, english = american]{csquotes}
\MakeOuterQuote{"}

\lstdefinestyle{log}{
    basicstyle=\ttfamily\small,   % Monospace font
    breaklines=true,              % Automatic line breaking
    frame=single,                 % Frame around the code
	breakindent=0px,
    backgroundcolor=\color{gray!10}, % Light gray background
    numbers=none,                  % No line numbers for logs
	keywordstyle=\color{red}, % Makes keywords red
    escapeinside={(*@}{@*)}, % Allows inline LaTeX within listings
	basicstyle=\scriptsize
}

\lstdefinestyle{bash}{
    language=bash,
    basicstyle=\ttfamily\small,   % Monospace font
    breaklines=true,              % Automatic line breaking
    frame=single,                 % Frame around the code
    backgroundcolor=\color{gray!10}, % Light gray background
    showstringspaces=false,        % Do not mark spaces in strings
	basicstyle=\scriptsize
}

\lstdefinestyle{xml}{
    language=XML,                     % Set language to XML
    basicstyle=\ttfamily\small,       % Monospace font
    frame=single,                     % Frame around the code
    breaklines=true,                  % Automatic line breaking
    backgroundcolor=\color{gray!10},  % Light gray background
    showstringspaces=false,           % Do not mark spaces
    columns=fullflexible,             % Proper alignment
    morekeywords={node, key, value},  % Add custom XML tags
    commentstyle=\color{green!50!black}, % Color for comments
	basicstyle=\scriptsize
}
\usepackage{graphicx}
\usepackage{subcaption}

\begin{document}
\title{Towards Improving Intrusion Detection Systems Using Capture the Flag Events}

\titlerunning{Towards Improving IDS Using CTF Events}
%\author{Manuel Kern\inst{1}\orcidID{0000-1111-2222-3333} \and
%Second Author\inst{2}\orcidID{1111-2222-3333-4444}}
\author{Manuel Kern\inst{1} \and Florian Skopik\inst{1} \and Max Landauer\inst{1} \and Edgar Weippl\inst{2}}
\authorrunning{M. Kern et al.}
\institute{Austrian Institute of Technology, Giefinggasse 4 1210 Vienna \and
University of Vienna, Universitaetsring 1, 1010 Vienna}
\maketitle              % typeset the header of the contribution
\begin{abstract}
In cybersecurity, Intrusion Detection Systems (IDS) serve as a vital defensive layer against adversarial threats. Accurate benchmarking is critical to evaluate and improve IDS effectiveness, yet traditional methodologies face limitations due to their reliance on previously known attack signatures and lack of creativity of automated tests. This paper introduces a novel approach to evaluating IDS through Capture the Flag (CTF) events, specifically designed to uncover weaknesses within IDS. CTFs, known for engaging a diverse community in tackling complex security challenges, offer a dynamic platform for this purpose. Our research investigates the effectiveness of using tailored CTF challenges to identify weaknesses in IDS by integrating them into live CTF competitions. This approach leverages the creativity and technical skills of the CTF community, enhancing both the benchmarking process and the participants' practical security skills. We present a methodology that supports the development of IDS-specific challenges, a scoring system that fosters learning and engagement, and the insights of running such a challenge in a real Jeopardy-style CTF event. Our findings highlight the potential of CTFs as a tool for IDS evaluation, demonstrating the ability to effectively expose vulnerabilities while also providing insights into necessary improvements for future implementations. 
\end{abstract}

\section{Introduction}
%ALT for 2 paragraphs: Intrusion detection systems (IDS) serve as a critical second layer of defense in cybersecurity. Despite extensive research \cite{landauer2024introducing,shiravi2012toward}, accurately benchmarking these systems remains challenging \cite{idsbenchmeth,ranum2001experiences,maseer2021benchmarking}. Traditional methods, such as AV-TEST \cite{avtest} and user reviews \cite{gartnerpeer}, rely on known attack signatures, lacking objective analysis. In contrast, Capture the Flag (CTF) competitions, with their dynamic engagement and community-driven challenges \cite{ctftime2024,pwn2own}, present a novel venue for revealing IDS weaknesses.
Intrusion detection systems (IDS) detect adversarial behavior as a crucial second layer of defense following preventive measures.
Over the years, developing and assessing these systems have been a central focus of cybersecurity research \cite{landauer2024introducing,shiravi2012toward}. Measuring their effectiveness poses significant challenges \cite{idsbenchmeth,ranum2001experiences,maseer2021benchmarking}. Expertise is required to benchmark these systems accurately, and even then, the process demands considerable effort. Traditional benchmarks, such as AV-TEST \cite{avtest}, rely on curated datasets of known malware, which may not reflect the complexity of real-world threats or evolving attack techniques. User reviews on platforms like Gartner Peer Reviews \cite{gartnerpeer} are, while providing anecdotal insights, subject to potential biases. These include variability in reviewer expertise, and the influence of vendor sponsorship or incentivized reviews, which can skew perceptions of IDS performance.
Similarly, attack simulation software using contemporary attack techniques \cite{landauer}, even with the advancements presented by MITRE ATT\&CK evaluations \cite{mitreeval} making results public and comparable, still depends heavily on knowledge of already known threats.
Given these limitations, penetration testing and red teaming exercises become invaluable. They simulate realistic attack scenarios aimed at bypassing IDS unnoticed %\cite{mitreattack}
. However, the high costs associated with professional red teaming and the confidentiality of the results, typically not disclosed due to proprietary interests, restrict their broader applicability. Furthermore, the outcome of these tests varies considerably based on the methodology and the testers' expertise. These factors highlight the need for %more objective and 
dynamic benchmarking methods. In contrast, Capture the Flag (CTF) competitions offer a dynamic and engaging platform. These events %annually draw many of participants to CTFtime \cite{ctftime2024}, a global CTF tracker, 
foster a community adept at tackling various security problems—from simple riddles to complex attack simulations. CTFs encourage innovative thinking and practical security skills enhancement, exemplified by competitions like Pwn2Own \cite{pwn2own}, which incentivize finding zero-day vulnerabilities that directly contribute to product security improvements.
To our knowledge, CTF challenges have not yet been used to find weaknesses in IDS. This makes CTF events a yet underexplored tool for evaluating IDS performance. Traditional technical tests often follow predefined patterns, overlooking the creativity and adaptability of real attackers. CTF participants, particularly white-box hackers, emulate adversaries’ dynamic behaviors by leveraging their skills for the thrill and challenge of problem-solving. This allows for more realistic and comprehensive testing of IDS, mirroring the tactics of real-world attackers. By integrating CTF challenges designed to expose IDS vulnerabilities, we harness the creativity and technical proficiency of the CTF community to advance IDS evaluation beyond static, signature-based approaches.

Our study explores the potential of CTF challenges to find weaknesses in IDS. We pose the research question: \textit{To what degree can Capture the Flag challenges improve the identification of weaknesses in intrusion detection systems?} We propose a novel approach that integrates CTF challenges explicitly designed to test and expose vulnerabilities in IDS setups. This integration aims to leverage the creativity and technical proficiency of the CTF community and the broad accessibility and engagement of CTF events. 
However, designing and integrating such challenges into CTF competitions and ensuring their relevance and effectiveness present unique challenges. 
We outline our contributions as follows: (i) We develop a tailored design methodology for IDS-focused CTF challenges; (ii) we establish a scoring system that motivates participants and enriches their learning experience, while at the same time incentivizes them to evade detection systems; (iii) we share empirical insights and practical lessons learned from executing a CTF challenge aimed at detecting IDS vulnerabilities. This approach not only broadens the evaluation landscape for IDS but also enhances the practical skills of cybersecurity practitioners in an engaging and competitive environment.

The structure of this paper is as follows: Section~\ref{sec:relatedwork} presents a review of related work and establishes the differentiation from existing approaches.
Section~\ref{sec:methodology} outlines our study's methodology, detailing the design, challenge format, delivery mechanisms, scoring system, and the proposed methodology for challenge creation. Section~\ref{sec:poc} demonstrates the application of our methods during a live CTF event, presenting initial findings. The results are further explored and discussed in Section~\ref{sec:discussion}, where we also address the research question of our study and provide future directions. Section~\ref{sec:conclusion} provides concluding remarks on our research.

\section{Background \& Related Work}
\label{sec:relatedwork}
%soa of ids benchmarking
%challenges specific to 
Benchmarking of intrusion detection systems (IDS) has been a focal point of research for several decades. Athanasiades et al. \cite{idsbenchmeth} critically reviewed existing IDS benchmarking tools, underscoring the necessity for new tools that better simulate realistic network activities and improve repeatability and scalability. Similarly, Ranum \cite{ranum2001experiences}  emphasized the importance of precise measurements for IDS effectiveness and provided guidelines for IDS benchmarking.
Recent work by Maseer et al. \cite{maseer2021benchmarking} involved evaluating the performance of supervised machine learning algorithms on 31 anomaly-based IDS solutions, utilizing metrics such as accurate positive and negative rates, accuracy, precision, recall, and F-Score. This approach is seen in many other papers benchmarking IDS. There has been significant effort in developing datasets to benchmark newly proposed detection algorithms, for example, seen in works by Landauer et al. \cite{landauer2024introducing} and Shiravi et al. \cite{shiravi2012toward}. 
Platforms like AV-TEST \cite{avtest} benchmark detection rates of commercial anti-malware/endpoint detection and response solutions. Moreover, recent years have seen the development of Adversarial Emulation Platforms aiming to set a baseline for IDS detection  \cite{landauer}. A notable initiative by MITRE called "evaluations" \cite{mitreeval} involves making benchmarks of IDS 
%using these tools \cite{caldera} 
public. However, Shen et al. \cite{Xiangmin24} criticize these benchmarks for lacking in-depth analytical insights. We argue that all these approaches rely on already known attack techniques and signatures.
While there are various Capture the Flag (CTF) formats \cite{zafar24,defcon2003,ictf,davis14,ccdcoe,bock18,ctftime2024,hackasat,pwn2own,kucek20}, %ccdc,ccdcoels24
including attack-defense or defend-only, which we will discuss in Sect. \ref{sec:chalformat}, they typically do not focus on identifying vulnerabilities within IDS. 
%-------------------------------------------------------------------------------
\section{Methodology}
\label{sec:methodology}
%-------------------------------------------------------------------------------
% Approach: Describe the theoretical framework, algorithms, or models used in the research.
% Implementation Details: Provide specific details about the tools, technologies, and data sets used.
% Experimental Design: If applicable, describe the experimental setup, including hardware and software configurations, testing environments, and evaluation metrics.

\begin{figure}[!b]
    \centering
    \includegraphics[width=.8\textwidth]{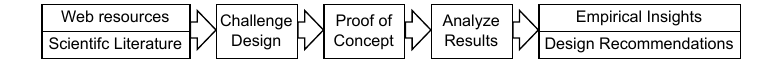}
    \caption{Overview of research methodology.}
    \label{fig:methodology}
\end{figure}
%\vspace{-1em}

We developed a structured methodology to verify if CTF challenges can be used to find IDS weaknesses. This approach integrates theoretical CTF challenge design principles with empirical testing and analysis. As illustrated in Figure~\ref{fig:methodology}, our process begins with the design of a CTF challenge aimed explicitly at finding weak spots in intrusion detection systems utilizing literature. We then deploy this challenge in a real-world CTF event setting. Observations from the event offer valuable data on the practicality of our method, including insights into participant strategies and behaviors. We conclude our study by discussing our empirical findings, enriched with feedback from participants and organizers. We also provide design recommendations and propose ideas for future research.

\subsection{Challenge Design}
The following section explores the design of a CTF challenge specifically for finding weaknesses in IDS configurations.

%1012  curl -o ctftime_events_2024.xml https://ctftime.org/event/list/archive/rss/
%1013  grep -o "<item>" ctftime_events_2024.xml | wc -l
%1014  grep -B 2 "Jeopardy" ctftime_events_2024.xml | grep -o "<item>" | wc -l
\subsubsection{Challenge Format} \label{sec:chalformat}
There are many CTF events running each year. To date the platform CTFtime \cite{ctftime2024} (one of the largest public databases about ongoing and future CTF events) counts 100 events for 2024.
By far, the biggest category is "Jeopardy," with 86 events, followed by "Attack-Defence" (further called Attack-Defense), with eleven events that happened in 2024.
\textbf{Jeopardy} competitions present participants with challenges that test their knowledge and skills in computer networking, security, and related domains. Each challenge requires problem-solving to uncover a hidden "flag," which proves task completion. \textbf{Attack-Defense} competitions involve dynamic interactions, emphasizing both offensive and defensive cybersecurity. Teams are tasked with securing their systems while simultaneously attempting to exploit vulnerabilities in the systems of their opponents. 
%\textbf{Hack-quest} on CTFtime \cite{ctftime2024} is a term used for various formats associated with cybersecurity challenges that focus on problem-solving, e.g. "Hack-A-Sat" \cite{hackasat}, a CTF that is designed to hack using phyiscal flatsat hardware and digital twins in space.
One of the biggest Attack-Defense challenges first held in 1996 \cite{zafar24} is hosted by Defcon \cite{defcon2003}. Another example of large-scale attack-defense is iCTF \cite{ictf} or Crossed Swords \cite{ccdcoe}%\cite{ccdcoecs24} 
including NATO states.
Other types are Defend-Only CTFs \cite{davis14} with a single third-party team playing an offensive or "red team" role. Teams are earning points by defending the infrastructure. An example of this format is the NATO CCDCOE Locked Shields exercise \cite{ccdcoe}.%\cite{ccdcoels24}.
%Well-known events are CCDC \cite{ccdc} or Locked Shields \cite{ccdcoels24}. 
King of the Hill \cite{bock18} where users have to take control of a target server, plant their identifier, and a scoring engine that monitors the target awards a team with points. 

Attack-Defense and Defend-Only competitions are highly dynamic, driven by participants' offensive and defensive actions. This variability creates an unpredictable environment, making it difficult to isolate and evaluate IDS consistently. For meaningful tests, a controlled and reproducible setup is essential.
Additionally, such high-dynamic environments prioritize securing infrastructure and exploiting opponents' vulnerabilities, where IDS functionality is only one component of many. In contrast, IDS benchmarking requires tasks specifically designed to evaluate detection accuracy, better suited to the structured Jeopardy format.
Reproducing scenarios of such high-dynamic environments is also complex. Real-time team interactions make it nearly impossible to recreate identical conditions for fair comparison. In contrast, Jeopardy challenges provide a structured, consistent environment, enabling more accurate assessments of IDS detection rates without interference from unrelated variables.
In summary, the Jeopardy format provides a more controlled and task-oriented environment where challenges can be designed to precisely evaluate the capabilities of IDS systems. Identical tasks and attack scenarios can be presented to participants, ensuring consistent conditions. Participants can focus on understanding and leveraging the weaknesses of the IDS systems to attack. 
%Furthermore, the static nature of Jeopardy challenges reduces the complexity of the environment, making it easier to isolate the IDS's accuracy from other variables.
In our prototype implementation, we investigated further if the Jeopardy format is suitable for finding IDS weaknesses.

\subsubsection{Challenge Delivery}
%ALTGPT CTF challenges are typically hosted on shared infrastructure managed by event organizers. While challenges like reverse engineering are easily hosted with simple file downloads for offline analysis, others require more sophisticated setups. Maggioni et al. \cite{niccol24} found that containerized deployments using orchestrators like Kubernetes provide particular value for complex scenarios. For challenges that test Intrusion Detection Systems (IDS), ensuring no cross-team interference and maintaining team isolation is critical, as IDS monitor attack patterns and behaviors that must be attributed uniquely to each team. Shared setups, therefore, pose challenges in distinguishing participant activities without complex workarounds. Hosting on a participant’s local machine, like using virtual machines, is not suitable for hacking challenges due to risks of environment tampering and challenges in external verification of flag retrieval. Consequently, the ideal deployment model is a per-team hosted infrastructure. This model prevents cross-team interference, maintains centralized control, and ensures the integrity of the IDS benchmark. Whether deploying on dedicated servers, containers, or using orchestration depends on the specific requirements of the IDS. Key factors in choosing the right CTF environment include scalability, isolation, and resource management, necessitating a process manager to handle multiple simultaneous processes such as web servers and application servers alongside the IDS.

Capture The Flag (CTF) challenges are typically hosted on shared infrastructure by the event organizers. While some challenge types, such as reverse engineering (e.g., analyzing a binary), are straightforward to host because they only require file downloads for offline analysis, others demand more complex infrastructure.
Maggioni et al. \cite{niccol24} evaluated a range of hosting scenarios, from dedicated servers to virtualization, containerization, and complex orchestration frameworks. Their findings highlight the particular value of containerized deployments, whether using simple orchestrators, task runners, or more advanced solutions like Kubernetes.
% This is mainly because of easier deployment, configuration changes and scalability.
When implementing a challenge to evaluate weaknesses in IDS, team isolation is particularly important. IDS monitor for attack patterns and abnormal behavior. To ensure that the results are comparable and not influenced by interference between teams these attack patterns and behaviors must be uniquely attributed to individuals.
When using shared infrastructures, such as only one installation for all teams, clear distinctions between participant activities and ensuring no interference between teams require complex workarounds. 
%Although one approach could be reserving unique IP address ranges or assigning specific host header values at the network level per team. However, attacks targeting the host itself may not be easily distinguished by such methods, also teams might interfere other teams on purpose. As a result, shared infrastructure introduces significant limitations in realism for IDS benchmarking, as certain interactions and attack scenarios cannot be adequately isolated.
Hosting challenges on a participant's local machine risks compromising the integrity and fairness of the event, as players can tamper with the environment where the flag is stored. Consequently, offline installations, like virtual machines, are unsuitable for hacking challenges where external verification of flag retrieval is critical.
%Hosting an offline installation, such as a virtual machine deployed on the player's machine, poses significant challenges. In reverse engineering, stenography, or cryptography challenges, retrieving the flag is an integral part of the task and is hidden within the binary or data provided. However, in hacking challenges, the flag is typically stored on the infrastructure itself.
%An external program must verify whether the participant has successfully retrieved the flag. This verification process must ensure that no external manipulations or detections interfere with the intended flow of the challenge. In offline installations, maintaining this integrity can be violated because the player has full control over the environment. Consequently, offline installations are not suitable, as they risk compromising the challenge's integrity and fairness.
Given these constraints, the ideal deployment model is a hosted, per-team infrastructure. This approach eliminates the risk of cross-team interference, ensures that the environment remains under centralized control, and preserves the integrity of the IDS benchmark.
Regarding the hosting scenario, whether a dedicated server, virtual machine, containerization, or complex orchestration is used for each team depends on the IDS's requirements. The key factors of such a challenge in CTF environments are scalability, isolation, and resource management.
Since multiple processes typically need to run (e.g., a web server, an application server, and an IDS), a process manager is required.
% (e.g., Supervisord in Docker environments). 
% Alternatively, service startup can be managed across multiple containers. The specific design choice depends on the challenge's specific implementation and technical requirements.
\subsubsection{Scoring System}
\label{sec:scoring}
It is essential to keep participants engaged during a Jeopardy CTF to test an IDS's detection weaknesses.
In an online competition, this is primarily achieved through the scoring system. The score motivates participants to solve challenges and find optimal solutions. 
%During a Jeopardy CTF event, participants are presented with several challenges, each with a different maximum score, depending on its difficulty.
Understanding the scoring system is crucial for success \cite{davis14}. 
% A well-designed scoring system significantly impacts both the enjoyment of the challenge and the incentives for participants to perform well \cite{davis14}.
Various scoring approaches are used in CTF's. There is no single "best" solution, and many events create their own scoring mechanisms.
Zafar et al. \cite{zafar24} explore the development and algorithms behind CTF scoring systems. Typically, Jeopardy-style CTF's employ either static or dynamic scoring, with rare instances of manual scoring by a jury.
In static scoring, the points awarded for solving a challenge remain fixed throughout the event and are usually based on the task's difficulty.
In contrast, dynamic scoring changes throughout the event. For example, the score for a challenge may decrease with each new submission or be adjusted based on the time elapsed since the challenge was released.
While static scoring is simple to explain, implement, and track, dynamic scoring aims to address some issues. In Capture the Flag (CTF) competitions, a common strategy known as \textbf{flag hoarding} involves participants concealing their actual scores until the last moment. This tactic can also mislead other teams about the difficulty of a challenge. If no solutions appear publicly, teams might wrongly assume that a challenge remains unsolved and is particularly challenging. Consequently, teams often withhold their flag submissions to maximize their scores, choosing to reveal them only at the very end of the competition. Different perceptions in \textbf{challenge difficulty} are typically addressed by dynamic scoring, based on how many teams solve a challenge; supporting scores reflect the true difficulty. As more participants achieve similar scores, tie-breakers become increasingly necessary. Dynamic scoring can address this by reducing the score for subsequent teams solving the challenge. More importantly, it also mitigates flag hoarding by penalizing late submissions of flags, discouraging teams from delaying their hand-ins until the final moments.
Dynamic scoring is typically implemented through a score decay formula.
%, which defines how scores decrease over time or with the number of submissions. A typical decay formula for CTF's could look as follows:
%\[
%a = \text{max points}, \quad b = \text{min points}, \quad s = \text{solve %threshold}.
%\]

%%FIX MINUS?
%\[
%f(x) = \frac{b - a}{s^2} x^2 + a
%\]

We design a scoring to support our goal of motivating participants to identify weaknesses in IDS configurations. In this context, our scoring model rewards teams that remain undetected during the challenge. To enhance the educational value and the enjoyment of the CTF event, we propose attributing points for stealthy approaches. For instance, teams earn points for triggering only a few low-severity alerts, such as generating application errors that do not directly signal an ongoing attack. This encourages participants to refine their evasion strategies through more subtle techniques.
We define two key properties: (i) teams earn points when submitting a flag, allowing them to gain credit even for intermediate results; (ii) teams that trigger fewer alerts from the IDS systems will score higher points.
The first property ensures that teams can still earn points for progress, even if they have not yet found the optimal solution. The second property incentivizes teams to find more refined solutions, such as evading detection by bypassing multiple IDS or operating under the radar.
To implement this, we base the scoring on the severity of the triggere dalerts. IDS systems typically assess detections according to their criticality. For example a 404 error thrown on a web application or a service restart is an informative event with a low score that could be the first indicator of an ongoing attack, but a malware signature is a high-severity alert that identifies a specific attack with high confidence. By summing up all alerts triggered until the target is reached, we can derive a score. This score forms the basis to guide points allocation.
%In Jeopardy-style CTF games, dynamic scoring functions often rely on decay mechanisms \cite{zafar24}. 
For our challenge, we propose using a convex decay function to decrease the points awarded based on higher alert scores. %While a static scoring function could suffice, our goal is to emphasize evasion techniques and make it a critical factor in the challenge. 
Achieving zero alerts should yield significantly more points than triggering some alerts. Therefore the scoring function uses a logarithmic decay with a steepness of 0.2. %, as illustrated in Figure~\ref{fig:logarithmic_decay}. 
A perfect solution triggers no alerts, yielding 500 points.
% This makes a convex function the better choice. An example implementation could look like this:
\[
a = \text{max points}, \quad b = \text{min points}, \quad s = \text{steepness}.
\]
\[
f(x) = max(b, a - (s * ln(x)) * (a - b))
\]

\begin{comment}
\begin{figure}[ht]
    \centering
    \includegraphics[width=0.4\textwidth]{logarithmic_decay_plot.pdf}
    \caption{Logarithmic decay function for increasing alert severity score, e.g., more high-confidence alerts triggered, results in a smaller score. Maximum points for the challenge are 500, decreasing to a minimum of 100 based on alerts triggered.}
    \label{fig:logarithmic_decay}
\end{figure}
\end{comment}

%In recent years, additional scoring features have been implemented to enhance CTF competitions \cite{zafar24,kucek20}. These include "first blood," where extra points are awarded for the first successful solve of a challenge, score deductions for using hints, and penalties based on the number of attempts. Some challenges also include multiple flags, further diversifying the scoring dynamics. Researching these features clearly exceeds this study.
%Further research is necessary to evaluate the impact of these features on player behavior and the overall effectiveness of CTF competitions for benchmarking purposes.

We also looked into scoring techniques typically used for Attack-Defense based CTFs. E.g., attack points, defense points, service level agreement (SLA) \cite{zafar24}, and King of the Hill \cite{bock18}. These focus on offensive tactics, e.g., capturing flags and controlling services. These models prioritize exploit speed or service availability rather than evasion or stealth.
The decay function, used with the sum of alert severity scores, on the other hand, provides a fine-grained scoring system and is easy to understand. Participants are not only incentivized to solve the challenge but also to do so with minimal IDS detection. This further reflects how well participants adapt their strategies to bypass detection.

% In conclusion, the final choice of scoring system depends on the specific technical implementation and the format of the CTF event. The application of attack-defense-based scoring systems should not be ruled out and warrants further investigation. However, in our demo setup, we found that a decay function based on severity scores was easy to implement, straightforward to understand, and effectively supported the desired properties.

\subsubsection{Challenge Design Implementation Requirements}
Technical implementation forms the core of any CTF challenge and requires significant effort from designers. A challenge should be realistic but not rigidly bound to realism, ensuring it remains engaging for participants \cite{davis14}. It fosters a player's understanding of the underlying concepts of the problem and follows a logical solution path, avoiding unnecessary guesswork \cite{cryptax}. 
%Many examples of challenges are available in archives such as the CTF Archive \cite{ctfarchive}.
Chung and Cohen \cite{chung14} emphasize that the success of a CTF event hinges on the quality of its challenges, which must strike a balance between difficulty and engagement. 
%Challenges should provide subtle hints to guide participants and foster a rewarding experience without undue frustration. Overly simplistic challenges, provide limited educational value and fail to engage participants meaningfully. Conversely, overly complex challenges risk alienating players. Challenges relying on brute-force techniques, such as password or file name guessing, are problematic. These approaches do not contribute to meaningful learning outcomes and should be avoided unless intentional, documented and manageable.
Our challenge leverages real-world software, IDS components, and infrastructure setups to achieve realism. 
%CTF competitions typically run for one to two days, during which many participants scarcely rest or sleep \cite{davis14}. 
%This limited timeframe reflects scheduling constraints, as participants can often commit to a weekend but not to extended durations. 
While this compressed schedule adds intensity and excitement to the event \cite{davis14}, it also makes certain scenarios, such as stealth attacks or social engineering, impractical. In contrast, sophisticated real-world attackers may execute carefully planned campaigns over weeks or months.
Further, a typical CTF challenge is not about finding zero days as noted by Cryptax \cite{cryptax}. There shall be a sample solution foreseen by the developers of the challenge. Defining a Jeopardy challenge without a known solution is not good etiquette and could disrupt the integrity of the CTF event when not clearly communicated. This is not ideal, since we want to benchmark our perfectly tuned IDS system with no known weaknesses.
However, the statement of Cryptax cannot be generalized, as demonstrated by the example of the Pwn2Own \cite{pwn2own} challenge. Participants search for zero-day exploits in this competition and are rewarded with significant prize money. 
%The format of Pwn2Own does not fall under the Jeopardy style. Instead the participants are challenged to find to find security bugs in widely used products. It was established in 2007 by Dragos Ruiu as a response to Apple \cite{pwn2own2007}. In cases where Apple security vulnerabilities were discovered, Apple MacBook Pros were awarded to the finders. Participants are able to prepare for the event since way longer than at a Jeopardy style CTF.
The time limitation of the Jeopardy format highlights a potential drawback. Nevertheless, we chose this format for our initial proof of concept to gain practical insights. By running this challenge, we aim to evaluate the feasibility of using CTFs to benchmark IDS systems and refine our approach based on observed outcomes in future research.

\subsection{Challenge Design Methodology}
We developed a four-step methodology for designing an evasion-based CTF challenge. % that is shown in Fig. \ref{fig:designchallenge}. 
%begin{figure}[ht]
%	\centering
%	\begin{tikzpicture}[node distance=1.4cm]
		% Nodes
%		\node (step1) [process] {1. Set the Scope};
%		\node (step2) [process, below of=step1] {2. Design the Infrastructure};
%		\node (step3) [process, below of=step2] {3. Establish Rules and Scoring};
%		\node (step4) [process, below of=step3] {4. Test and Validate};
		% Arrows
%		\draw [arrow] (step1) -- (step2);
%		\draw [arrow] (step2) -- (step3);
%		\draw [arrow] (step3) -- (step4);
%	\end{tikzpicture}
%	\caption{Proposed methodology for the challenge design.}
%	\label{fig:designchallenge}
%\end{figure}
First, we \textbf{set the scope}. We decide which IDS components to evaluate (for example, Wazuh \cite{wazuh} or ModSecurity \cite{modsecurity}) and which attack vectors to include (such as SQL injection or XSS). We also define metrics for our evaluation. In our case, we focus on the number of alerts. If no alerts are triggered but the challenge is solved, this indicates a potential false-negative scenario. Unfortunately, the details behind these false negatives remain unclear. While options like screen recording or deploying agents on participants' systems could provide more granular insights, these approaches may face resistance due to privacy concerns and could discourage participation. Thus, our methodology prioritizes non-intrusive means of gathering data.
Next, we \textbf{design the infrastructure}. We deploy target services (for example, web servers) behind one or more IDS components. We also integrate logging and monitoring to capture performance metrics and record attacks for further analysis.
It is crucial to \textbf{establish rules and scoring} for assigning points to successful exploits (see Sect.~\ref{sec:scoring} for details). We must specify rules, such as disallowing denial-of-service attacks \cite{ptacek08}, and give clear instructions to maintain a stable environment.
Lastly, we \textbf{test and validate} the challenge. We confirm that it functions as expected and can be completed within the allotted time. Ideally, beta testers provide feedback on design and difficulty. We then document how the solution process should unfold, including any expected alerts and metrics.

\subsection{Limitations}
\label{sec:limitations}
Our approach primarily focuses on identifying weaknesses in intrusion detection systems regarding the detection of attacks, specifically false negatives, where the system incorrectly classifies an event as benign when an attack is actually occurring. However, when benchmarking IDS systems, other factors such as false positives, where a benign event is incorrectly labeled as malicious, have to be considered. Further, our methodology does not account for other benchmarking attributes, such as the number of events an IDS can process or the processing time from event to alert. 
Although the scoring system seems dynamic, it is effectively static within the context of the game, as scoring applies static to the given challenge. For this reason, it retains its usual weaknesses, such as flag hoarding and inaccurate difficulty adjustment.
In a game setting where multiple challenges are hosted, dynamic scoring should ideally be implemented across challenges.

\section{Proof of Concept}
\label{sec:poc}
To explore the potential of CTF competitions for identifying vulnerabilities in IDSs, we partnered with the Austrian Cyber Security Challenge (ACSC) \cite{ACSC}. 
%This annual event draws diverse participants, from students to professionals, challenging them with network security, cryptography, and ethical hacking tasks. Early in our research, we introduced the idea of using a CTF to benchmark IDS systems to the ACSC organizers, who welcomed our concept and allowed us to contribute a challenge.
In this section, we describe how we designed and deployed a Jeopardy challenge at the ACSC and share insights into how participants perceived and solved it.

\subsection{Technical Setup}
\begin{figure}[htbp]
  \centering
  \begin{subfigure}[c]{0.47\textwidth}
    \centering
    \includegraphics[width=\textwidth]{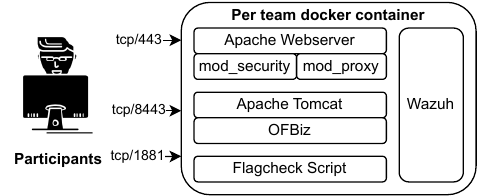}
    \caption{Challenge design.} %: Apache mod\_proxy, as well as Apache Tomcat point to the Apache OFBiz Tomcat Webapplication. Wazuh and mod\_security trigger alerts on intrusion attempts. A "FlagCheck" script is used to derive the final score.}
    \label{fig:challengedesign}
  \end{subfigure}
  \begin{subfigure}[c]{0.47\textwidth}
  \hfill % This command adds space between the two subfigures if needed
    \includegraphics[width=\textwidth]{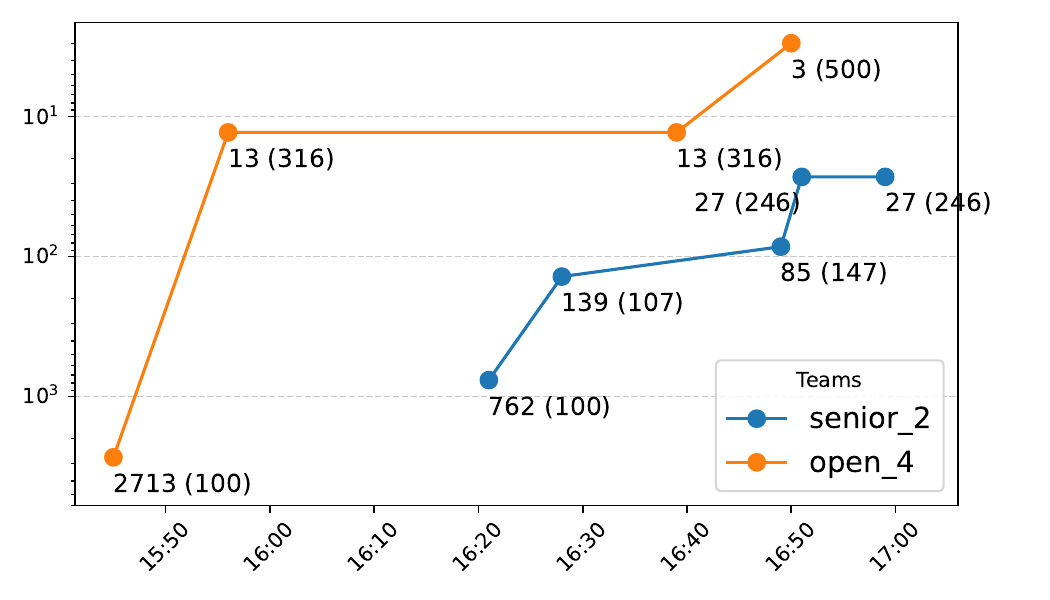}
    \caption{Challenge solves. %all solves recorded during the challenge day. % The total number of detection severity alerts for each submission is shown, with the achieved scores in brackets. The lowest possible detection score is three, as a single alert with a severity of three is triggered when Wazuh starts.
    }
    \label{fig:scoresovertime}
  \end{subfigure}
  \caption{Outline of design and challenges solves at the ACSC \cite{ACSC}.}
  \label{fig:test}
\end{figure}

We had some constraints to overcome. The infrastructure used by ACSC is Kubernetes. Therefore, we had to stick with containerization and Linux. The ACSC planned to host 16 teams, with each team up to 4 people, for eight hours daily. The event took place onsite in Vienna. %, with containers hosted in the Exoscale cloud \cite{exoscale}
Given the Kubernetes-based hosting, we focused on Linux-compatible IDS solutions. We excluded network intrusion detection systems (NIDS) like Suricata \cite{suricata} due to the complexities of running them in containers and instead selected Wazuh \cite{wazuh} and mod\_security \cite{modsecurity} with Apache for our setup. 
We wanted to shift the focus to evasion, so we selected a recent, realistic, yet public vulnerability for easy exploitation. We searched the Exploit Database \footnote{\url{https://www.exploit-db.com/}}% \cite{exploitdb} 
and chose CVE-2024-38856 \cite{CVE-2024-38856}
, a critical flaw in Apache OFBiz \cite{ofbiz}, an open-source enterprise resource planning (ERP) tool from the Apache Software Foundation. OFBiz versions up to 18.12.14 allow improper authorization and potential code execution. Public exploits were available, which made it easier for attackers to gain initial access. We combined Apache Webserver, Apache Tomcat, Wazuh, mod\_security (with the Core Rule Set \cite{coreruleset}), and OFBiz in a container. Fig. \ref{fig:challengedesign} shows the overall design. We then planted a flag in the Tomcat user's home directory.

\begin{comment}
\begin{figure}[ht]
    \centering
    \includegraphics[width=.8\textwidth]{ChallengeArchtitecture-v2.drawio-crop}
    \caption{Challenge for the ACSC \cite{ACSC}: Apache mod\_proxy, as well as Apache Tomcat point to the Apache OFBiz Tomcat Webapplication. Wazuh and mod\_security trigger alerts on intrusion attempts. A "FlagCheck" script is used to derive the final score.}
    \label{fig:challengedesign}
\end{figure}
\end{comment}

Intrusion attempts were monitored by Wazuh and mod\_security, triggering alerts. Participants' efforts to evade detection were scored based on the number and severity of alerts triggered, with the results determined by the custom "FlagCheck" script that validated the flag and aggregated alert severity. Listing \ref{lst:CheckScript} shows an example of the alerts triggered, which aggregated to a final score. A higher number of alerts reduced the final score.
When scanning the webserver, multiple alerts were triggered. But executing the exploit, triggered no alerts in the default IDS configuration.
When analyzed closely, we found that the vulnerability lies in the endpoint called "ProgramExport". We configured a Wazuh rule that checks if "ProgramExport" was called in the URL string and assigned a severity of twelve. 
%As a false flag, we referenced an older vulnerability, CVE-2023-51467 \cite{CVE-2023-51467}, which also targeted "ProgramExport" in the rule description. This reference became visible to participants during their analysis of the triggered alerts.
\begin{comment}
\begin{lstlisting}[style=xml, caption={Custom Wazuh rule configuration.}, label={lst:tomcat}]
<group name= "custom,tomcat">
<rule id="100002" level="12">
	<if_sid>31108</if_sid>
	<description>Possible execution of CVE-2023-51467: POST request to ProgramExport detected</description>
	<protocol>POST</protocol>
	<url>ProgramExport</url>
</rule>
</group>
\end{lstlisting}
\end{comment}
We encouraged participants to trigger fewer alerts to score higher using the scoring function described in Sect. \ref{sec:scoring}. Approaches that caused fewer alerts still earned a substantial score, while `loud' attacks earned fewer points. Listing \ref{lst:CheckScript} illustrates how the "FlagCheck" script returned which alerts were triggered so participants could refine their methods.
%When using the Apache Webserver (Port 443), the alert was triggered two times because of the proxy service in between.
\begin{comment}
\begin{lstlisting}[style=log, caption={Output of "FlagCheck" script with rule serverity score and flag in red.}, label={lst:CheckScript}]
telnet x.x.x.46 1881
Trying x.x.x.46...
Connected to x.x.x.46.
Escape character is '^]'.
Please input flag:
(*@\textcolor{red}{zeRIv2hmgSiaiaMm13SQf0VR}@*)
correct, calculating results
** Alert 1723752961.0: - ossec,pci_dss_10.6.1,gpg13_10.1,gdpr_IV_35.7.d,hipaa_164.312.b,nist_800_53_AU.6,tsc_CC7.2,tsc_CC7.3,
2024 Aug 15 20:16:01 e66d0e45ea51->wazuh-monitord
Rule: 502 (level 3) -> 'Wazuh server started.'
ossec: Manager started.

** Alert 1723753322.248: mail  - custom,tomcat
2024 Aug 15 20:22:02 e66d0e45ea51->/var/log/apache2/access.log
Rule: 100002 (level (*@\textcolor{red}{12}@*)) -> 'Possible execution of CVE-2023-51467: POST request to ProgramExport detected'
Src IP: 10.8.0.10
10.8.0.10 - - [15/Aug/2024:20:21:59 +0000] "POST /webtools/control/main/ProgramExport HTTP/1.1" 200 12099 "-" "-"

** Alert 1723753330.597: mail  - custom,tomcat
2024 Aug 15 20:22:10 e66d0e45ea51->/var/log/ids/ofbiz/access_log..2024-08-15
Rule: 100002 (level (*@\textcolor{red}{12}@*)) -> 'Possible execution of CVE-2023-51467: POST request to ProgramExport detected'
Src IP: 127.0.0.1
127.0.0.1 - - [15/Aug/2024:20:22:01 +0000] "POST /webtools/control/main/ProgramExport HTTP/1.1" 200 9757 "-" "-"

You had 3 alerts and a score of 27 (the lower the better ;)) ...
Connection closed by foreign host.
\end{lstlisting}
\end{comment}
\begin{lstlisting}[style=log, caption={Output of "FlagCheck" script with rule serverity score and flag (red).}, label={lst:CheckScript}]
Connected to x.x.x.46.
Escape character is '^]'.
Please input flag:
(*@\textcolor{red}{zeRIv2hmgSiaiaMm13SQf0VR}@*)
correct, calculating results
...
Rule: 100002 (level (*@\textcolor{red}{12}@*)) -> 'Possible execution of CVE-2023-51467: POST request to ProgramExport detected'
"POST /webtools/control/main/ProgramExport HTTP/1.1" 200 12099 "-" "-"
...
You had 3 alerts and a score of 27 (the lower the better ;)) ...
\end{lstlisting}
\subsubsection{Challenge Outline}
The challenge followed a straightforward sequence. Participants discovered two open ports and scanned for web directories, which triggered numerous alerts. 
In the first iteration of the challenge, attackers would generate alerts before discovering the OFBiz endpoint.
After finding Apache OFBiz and identifying its specific version, they would find a public exploit to achieve code execution. This would allow them to find and access the flag in the Apache OFBiz user's home directory. Next, they would submit that flag to a dedicated check script, which shows them the number of alerts triggered and their resulting score. Equipped with this feedback, they would refine their techniques to reduce further alerts on subsequent attempts.
We also developed a "perfect" solution. While experimenting, we discovered that Apache Tomcat did not write logs to disk if the connection stayed open for a long period. Because of that missing log entry, Wazuh never triggered an alert. This issue only appeared when the connection went directly to Tomcat. It did not occur if requests were routed through Apache, as the Apache Webserver is not facing the same issue. This behavior indicates an underlying security problem. We are still investigating why Apache Tomcat behaves that way.
%Participants could reset the challenge at any time using the "FlagCheck" script. 
This resulted in a fresh setup of the challenge, which was also triggered when the flag was posted. Participants could also reset the challenge anytime by invoking the "FlagCheck" script. 
\subsubsection{Testing and Validation}
Two junior penetration testers were given the challenge with a maximum timeframe of two hours. While they quickly solved the first part of the challenge within this timeframe, they needed additional time to fully understand the core objective: adjusting their strategy to evade detection. Despite their efforts, they could not achieve a perfectly stealthy intrusion that triggers no alerts at all and and ultimately received a medium score, corresponding to the two detections shown in Listing~\ref{lst:CheckScript}.

Based on their feedback, we revised and clarified the challenge instructions, which we also provided in our GitHub repository \cite{challengepublic}.% as shown in Listing~\ref{lst:explain}. 
A security expert was also consulted to ensure the challenge had no unintended solutions and that the "FlagCheck" script was secure against tampering or exploitation.
To facilitate further research on evasion techniques, we implemented a routine to collect logs each time a flag was submitted or the container was reset. These logs were included in the data package accompanying the challenge. Participants were also required to submit write-ups detailing their strategies for bypassing detection before their final score was awarded. We provide an analysis of these submissions in the following subsection.
The complete challenge, including all configurations and the final solution, has been open-sourced and is available at GitHub \cite{challengepublic}.

\begin{figure*}[tb]
    \centering
    \includegraphics[width=1\textwidth]{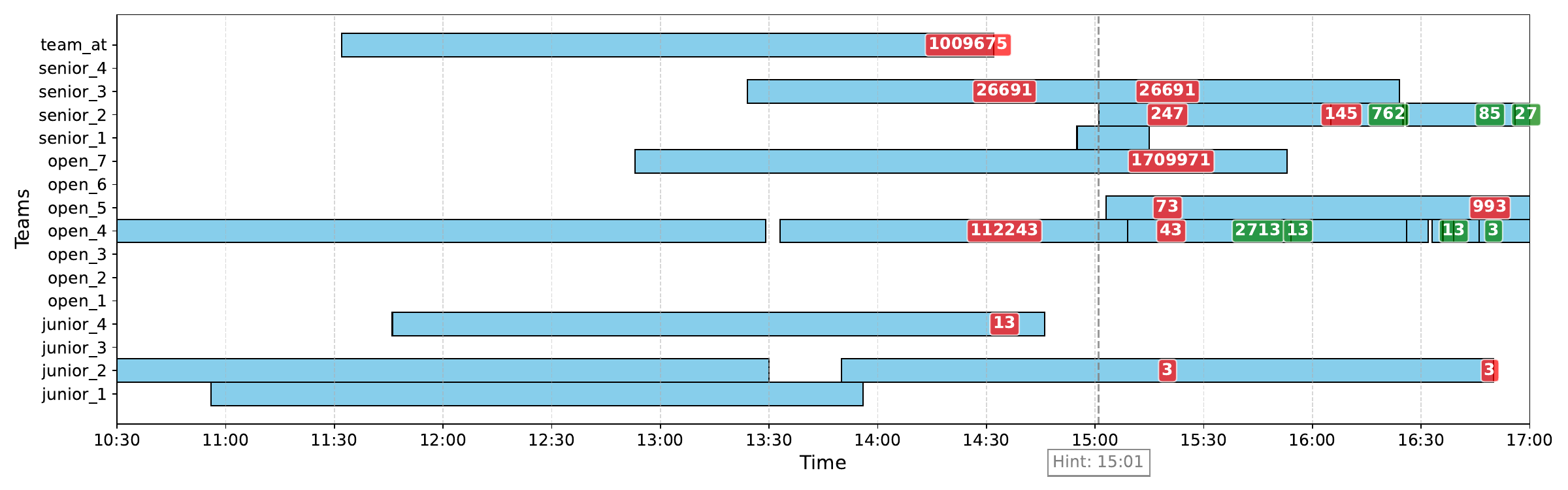}
    \caption{Light blue segments represent each container’s runtime. Red segments show alerts captured during manual log downloads, indicating unsolved challenges. Green segments depict scores for correct “FlagCheck” script submissions. %The light blue segments indicate the runtime of each container. Red segments represent the total number of alerts captured during manual log downloads, reflecting attempts where the challenge remained unsolved. Green segments show the scores awarded after participants successfully submitted the correct flag via the "FlagCheck" script. All participant submissions and collected log files are available in our data repository. Resets triggered by the "FlagCheck" script are not explicitly visible in the blue segments.
    }
    \label{fig:challengehosting}
\end{figure*}

\subsection{Event Day}
We provided the event organizers access to our challenge repository and deployed the challenge on their infrastructure. The final configuration on the CTFd platform \cite{ctfd}, including the challenge description is made public in our GitHub repository \cite{challengepublic}. %, is presented in Listing~\ref{lst:explain}.
Due to infrastructure constraints, all containers shared a single IP address. To address this, the organizers implemented a transparent proxy, which also removed access to the external Apache Tomcat port. This adjustment made the proposed "perfect solution" unattainable, as the only accessible port was the Apache Web Server. Consequently, at least one "ProgramExport" alert was consistently triggered even when holding the connection open. This unexpected change affected the participants' ability to use the same stealth techniques as the ideal solution, assuming the participants would not find new, undiscovered solutions.
Additionally, containers were configured to terminate automatically after three hours of runtime. Participants could also stop and restart containers through the CTFd interface. However, during these stops, our log-collection routine was not triggered, leading to gaps in the data collected for post-event analysis. This limitation restricted the completeness of our dataset for future investigation.
On the event day, we traveled to the venue to provide onsite support in case of any unexpected issues. Challenges were released in waves throughout the day, with our challenge included in the first wave and remaining available until the event concluded. Overall, our challenge was accessible for eight hours.
Multiple teams engaged with the challenge during this time. As shown in Figure~\ref{fig:challengehosting}, the first participants began working on the challenge at 10:29. Participants initiated containers hosting the challenge via a web platform. They were automatically terminated after three hours of runtime. Participants could also restart their containers through the web interface or the "CheckFlag" script. %Figure~\ref{fig:challengehosting} illustrates the Docker runtimes for each instance.

\begin{comment}
\begin{figure}[!bt]
    \centering
    \includegraphics[width=.8\textwidth]{solutions_scores_over_time.pdf}
    \caption{All solves recorded during the challenge day. The total number of detection severity alerts for each submission is shown, with the achieved scores in brackets. The lowest possible detection score is three, as a single alert with a severity of three is triggered when Wazuh starts.}
    \label{fig:scoresovertime}
\end{figure}
\end{comment}

\subsubsection{Observations}
In the afternoon, we initiated the download of log files from the active challenge containers to monitor participant progress. A significant observation was that many participants struggled to locate the correct entry point to the challenge. The initial phase of the challenge involved a classic enumeration task, closely resembling real-world penetration testing engagements \cite{pentestopensource}. 
Participants were expected to scan the web server for valid web applications, a process that inherently generates substantial noise. This is reflected in the high alert scores recorded during these attempts. The scores were calculated as the sum of all triggered alert severity scores (see also Listing~\ref{lst:CheckScript} for alert severity levels).
Unfortunately, most participants failed to identify the correct entry point for an extended period. The first team to discover the OFBiz directory, "open\_4," accomplished this at 13:38 after extensive scanning. Teams that relied heavily on web server scans triggered numerous alerts, including 400 and 404 errors. These errors, with a severity score of 5 each, quickly inflated the total detection scores into the thousands.
Since most participants struggled to locate the entry point, we released a hint to assist them, indicating Apache OFBiz as target. %, as shown in Figure~\ref{fig:hint}.
Shortly after the hint was released, the teams "senior\_2" and "open\_5" began engaging with the challenge. Team "senior\_2" was able to solve the challenge with a medium score, as shown in Listing~\ref{lst:CheckScript}, similar to the junior penetration testers in our test run.
In the end, only two teams successfully solved the challenge, despite 10 out of 16 teams starting it. Team "open\_4" achieved the perfect score minutes before the event concluded.
\begin{comment}
\begin{figure*}[h]
    \centering
    \includegraphics[width=1.0\textwidth]{hint.png}
    \caption{Hint that was released on the communication platform of the CTF.}
    \label{fig:hint}
\end{figure*}
\end{comment}
\subsubsection{Challenge Solves}
Fig. \ref{fig:scoresovertime} shows the total number of detection severity alerts per submission with corresponding scores in brackets on the x-axis. More alerts result in lower challenge scores. The minimum alert score for the challenge was 3, triggered by a single alert generated during the Wazuh initialization process. We received four challenge write-ups. 
\textbf{"open\_4" 15:56 - 13 detection points:} Identified a vulnerable version of Apache OFBiz and chose to exploit a lesser-known vulnerability, CVE-2024-45195 \cite{CVE-2024-45195}, with no ready exploit but a blog post. This approach avoided detection rules at the “ProgramExport” endpoint. Still, HTTP 404 errors from the empty web root triggered rules. \textbf{"open\_4" 16:50 - 3 detection points:} Refined solution after analyzing logs; avoided web root and directly executed payload. Achieved minimal detection score (3), earning maximum points (500).
\textbf{"senior\_2" 16:28 - 139 detection points:} The team used CVE-2024-38856 as intended, triggering the “ProgramExport” alert with their direct approach and not using evasion techniques, which led to a high detection score. \textbf{"senior\_2" 16:51 - 27 detection points:} In their second attempt, they employed basic evasion by URL-encoding some characters in “ProgramExport” and managed to execute their payload with fewer alerts, but not enough to fully evade detection, resulting in a medium detection score.

\section{Discussion} \label{sec:discussion}
\begin{table}[b!]
\vspace{-5mm}
\caption{Summary of observations and recommendations.} \label{tab:sum}
\centering
\scriptsize
\renewcommand{\arraystretch}{1.2} % Adjust row spacing
\setlength{\tabcolsep}{8pt} % Adjust column spacing
\begin{tabular}{|p{0.4\textwidth}|p{0.5\textwidth}|}
\hline
\textbf{Observation}                                                                                     & \textbf{Recommendation}                                                                                          \\ \hline
The enumeration phase was time-consuming and led to early abandonment by participants.                                  & Simplify or reduce enumeration requirements to allow participants to focus on IDS evasion techniques.       \\ \hline
Event infrastructure limitations affected logging and analysis. Changes introduced during hand-off to organizers caused inconsistencies.    & Use dedicated environments to ensure consistent setups and robust data capture. When relying on provided infrastructure, conduct pre-event tests to identify and mitigate potential issues. \\ \hline
Participants prioritized challenges they were familiar with due to time constraints.                                & Design shorter, modular tasks to keep participants engaged. \\ \hline
Participants were confused by the penetration testing format, which was unfamiliar in Jeopardy-style CTFs.     & Clearly define the challenge format and set appropriate expectations. Conduct further research into evasion-focused CTF formats.                       \\ \hline
\end{tabular}
\label{tab:discussion}
\end{table}
Our scoring system rewarded stealth by giving higher scores to participants who triggered fewer alerts. The implemented feedback mechanism enabled participants to iteratively refine their techniques, fostering a learning environment that supported the discovery of weaknesses in IDS. The requirement for write-ups documenting evasion strategies provided actionable insights, allowing us to identify their methods and refine our IDS.
One team discovered a vulnerability, which was unknown at the time of challenge creation (CVE-2024-45195), to bypass our detection rules, achieving a perfect solution. This demonstrates the potential of identifying weaknesses in IDS using CTF challenges, because of weak, static configuration of our IDS ruleset.
However, only a few teams engaged seriously with the challenge, and without the released hint, only one team would likely have solved it. This highlights the high level of expertise required and potential design or accessibility issues. 
After-event feedback identified barriers: long descriptions, initial 404 errors, and the time-intensive enumeration phase discouraged participants. The penetration testing-style challenge confused participants accustomed to typical Jeopardy-style tasks. 
Another participant of the CTF mistakenly assumed the challenge involved Windows Endpoint Detect and Response solutions and avoided it entirely. The eight-hour event format significantly stressed participants, making it difficult to engage with complex challenges and implement stealthy techniques, as reflected in a participant's blog post \cite{haunschmidt} and supported by Davis et al. \cite{davis14}. We've outlined our observations in Tab. \ref{tab:sum}

While Jeopardy-style CTFs offer a structured and isolated environment, ensuring sufficient time and focus for IDS benchmarking remains a challenge. Participants often prioritize speed due to event constraints, limiting the depth of stealth strategies they can employ compared to real-world attackers.

%-------------------------------------------------------------------------------
\section{Conclusion} \label{sec:conclusion}
%-------------------------------------------------------------------------------
% Summary: Summarize the main findings and contributions.
% Future Work: Suggest areas for future research or next steps to build on the work presented.
Our experiment demonstrated that Capture the Flag (CTF) events can serve as a valuable tool for identifying weaknesses in Intrusion Detection Systems (IDS). The controlled, reproducible nature of Jeopardy-style challenges allowed us to design a challenge focused explicitly on IDS evasion and detection weaknesses. However, several critical considerations emerged that highlight both the strengths and limitations of this approach. By reducing barriers, refining the challenge format, and addressing infrastructure constraints, future iterations can provide a more engaging and informative platform for evaluating IDS.

We have outlined design patterns for future iterations. Future research could also evaluate the possibilities of evaluating the accuracy by integrating false positives measurement, e.g. by injecting benign traffic. Also other benchmark metrics, e.g. time to alert or using  monitoring agents on the participants computer, could be evaluated.
Combining elements of Jeopardy and Attack-Defense formats could foster more focused IDS evaluations. Incorporating dynamic elements, such as timed rule updates or changing attack surfaces, could better simulate real-world conditions.

\begin{credits}
\subsubsection{\ackname} %A bold run-in heading in small font size at the end of the paper is
%used for general acknowledgments, for example: This study was funded
%by X (grant number Y).
Special thanks go to the Austrian Cyber Security Challenge organizers, particularly Manuel Reinsperger, Marco Squarcina, and Joe Pichlmayer, for their invaluable support with ideas and integrating the challenge. 

\subsubsection{\discintname}
The authors have no competing interests to declare that are relevant to the content of this article.
%It is now necessary to declare any competing interests or to specifically
%state that the authors have no competing interests. Please place the
%statement with a bold run-in heading in small font size beneath the
%(optional) acknowledgments\footnote{If EquinOCS, our proceedings submission
%system, is used, then the disclaimer can be provided directly in the system.},
%for example: The authors have no competing interests to declare that are
%relevant to the content of this article. Or: Author A has received research
%grants from Company W. Author B has received a speaker honorarium from
%Company X and owns stock in Company Y. Author C is a member of committee Z.
\end{credits}

%\begin{credits}
%\subsubsection{\ackname} A bold run-in heading in small font size at the end of the paper is
%used for general acknowledgments, for example: This study was funded
%by X (grant number Y).

%\subsubsection{\discintname}
%It is now necessary to declare any competing interests or to specifically
%state that the authors have no competing interests. Please place the
%statement with a bold run-in heading in small font size beneath the
%(optional) acknowledgments\footnote{If EquinOCS, our proceedings submission
%system, is used, then the disclaimer can be provided directly in the system.},
%for example: The authors have no competing interests to declare that are
%relevant to the content of this article. Or: Author A has received research
%grants from Company W. Author B has received a speaker honorarium from
%Company X and owns stock in Company Y. Author C is a member of committee Z.
%\end{credits}
\bibliographystyle{splncs04}
\bibliography{ifip-stealthctf}

\begin{thebibliography}{10}
\providecommand{\url}[1]{\texttt{#1}}
\providecommand{\urlprefix}{URL }
\providecommand{\doi}[1]{https://doi.org/#1}

\bibitem{ofbiz}
{Apache}: The apache ofbiz project. \url{https://ofbiz.apache.org/}, accessed:
  2024-12-27

\bibitem{idsbenchmeth}
Athanasiades, N., Abler, R., Levine, J., Owen, H., Riley, G.: Intrusion
  detection testing and benchmarking methodologies. In: Proc. of the First IEEE
  Int. Workshop on Information Assurance, 2003. pp. 63--72 (2003)

\bibitem{ACSC}
Austria, C.S.: Challenge. \url{https://verbotengut.at/} (2024), accessed:
  2024-12-27

\bibitem{avtest}
{AV-TEST Institute}: Antivirus testing procedures.
  \url{https://www.av-test.org/en/about-the-institute/test-procedures/},
  accessed: 2024-12-27

\bibitem{bock18}
Bock, K., Hughey, G., Levin, D.: King of the hill: A novel cybersecurity
  competition for teaching penetration testing. In: 2018 USENIX Workshop on
  Advances in Security Education. USENIX Association, Baltimore, MD (Aug 2018)

\bibitem{ccdcoe}
{CCDCOE}: Excercises. \url{https://ccdcoe.org/exercises/}, accessed: 2024-12-27

\bibitem{chung14}
Chung, K., Cohen, J.: Learning obstacles in the capture the flag model. In:
  USENIX 3GSE. USENIX Association, San Diego, CA (Aug 2014)

\bibitem{defcon2003}
Cowan, C., Arnold, S., Beattie, S., Wright, C., Viega, J.: Defcon capture the
  flag: defending vulnerable code from intense attack. In: Proceedings DARPA
  Information Survivability Conference and Exposition. vol.~1, pp. 120--129
  vol.1 (2003)

\bibitem{cryptax}
{Cryptax}: What makes a good ctf challenge?
  \url{https://cryptax.medium.com/what-makes-a-good-ctf-challenge-7bf4bf4fa112},
  accessed: 2024-12-27

\bibitem{ctfd}
{CTFd}: The easiest capture the flag platform. \url{https://ctfd.io/},
  accessed: 2024-12-27

\bibitem{ctftime2024}
{CTFtime}: All about ctf. \url{https://ctftime.org/}, accessed: 2024-12-27

\bibitem{davis14}
Davis, A., Leek, T., Zhivich, M., Gwinnup, K., Leonard, W.: The fun and future
  of {CTF}. In: USENIX 3GSE. USENIX Association, San Diego, CA (Aug 2014)

\bibitem{pentestopensource}
Faircloth, J.: Penetration Tester's Open Source Toolkit, Fourth Edition.
  Syngress Publishing, 4th edn. (2016)

\bibitem{gartnerpeer}
{Gartner Peer Insights}: Best siem.
  \url{https://www.gartner.com/reviews/market/security-information-event-management},
  accessed: 2024-12-27

\bibitem{hackasat}
{Hack-A-Sat}: World's first ctf in space. \url{https://hackasat.com/},
  accessed: 2024-12-27

\bibitem{ictf}
{iCTF}: The international capture the flag competition.
  \url{https://ictf.cs.ucsb.edu/}, accessed: 2024-12-27

\bibitem{kucek20}
Kucek, S., Leitner, M.: An empirical survey of functions and configurations of
  open-source capture the flag (ctf) environments. Journal of Network and
  Computer Applications  \textbf{151},  102470 (2020)

\bibitem{landauer}
Landauer, M., Mayer, K., Skopik, F., Wurzenberger, M., Kern, M.: Red team
  redemption: Open-source tools for adversary emulation (2024),
  \url{https://arxiv.org/abs/2408.15645}

\bibitem{landauer2024introducing}
Landauer, M., Skopik, F., Wurzenberger, M.: Introducing a new alert data set
  for multi-step attack analysis. In: Proc. of the 17th CyberSec
  Experimentation and Test Workshop. pp. 41--53 (2024)

\bibitem{niccol24}
Maggioni, N., Galletta, L.: A comparison of hosting techniques for online
  cybersecurity competitions. In: Intelligent Techn. for Interactive
  Entertainment (2024)

\bibitem{haunschmidt}
{Martin Haunschmid}: Acsc recap.
  \url{https://martinhaunschmid.substack.com/p/acsc-recap-und-ein-paar-ivanti-lucken},
  accessed: 2024-12-27

\bibitem{maseer2021benchmarking}
Maseer, Z.K., Yusof, R., Bahaman, N., Mostafa, S.A., Foozy, C.F.M.:
  Benchmarking ml for anomaly-based ids in cicids2017. IEEE Access  \textbf{9},
   22351--22370 (2021)

\bibitem{mitreeval}
{MITRE}: Att\&ck evaluations. \url{https://attackevals.mitre-engenuity.org/},
  accessed: 2024-12-27

\bibitem{CVE-2024-38856}
{NIST}: Cve - incorrect authorization vulnerability in apache ofbiz.
  \url{https://nvd.nist.gov/vuln/detail/cve-2024-38856}, accessed: 2024-12-27

\bibitem{CVE-2024-45195}
{NIST}: cve-2024-45195. \url{https://nvd.nist.gov/vuln/detail/cve-2024-45195},
  accessed: 2024-12-27

\bibitem{coreruleset}
{OWASP}: Crs project. \url{https://coreruleset.org/}, accessed: 2024-12-27

\bibitem{modsecurity}
{OWASP}: Modsecurity project. \url{https://modsecurity.org/}, accessed:
  2024-12-27

\bibitem{ptacek08}
Ptacek, T.H., Newsham, T.N.: Insertion, evasion, and denial of service: Eluding
  network intrusion detection

\bibitem{ranum2001experiences}
Ranum, M.J.: Experiences benchmarking intrusion detection systems. NFR Security
  White Paper  (2001)

\bibitem{Xiangmin24}
Shen, X., Li, Z., Burleigh, G., Wang, L., Chen, Y.: Decoding the mitre
  engenuity att\&ck enterprise evaluation. In: Proc. of the 19th ACM Asia CCS.
  p. 96–111. ASIA CCS '24, ACM (2024)

\bibitem{shiravi2012toward}
Shiravi, A., Shiravi, H., Tavallaee, M., Ghorbani, A.A.: Systematic approach to
  generate benchmark datasets for ids. Comput. Secur.  \textbf{31}(3),
  357--374 (2012)

\bibitem{suricata}
{Suricata}: Suricata ids/ips. \url{https://suricata.io/}, accessed: 2024-12-27

\bibitem{challengepublic}
{vexs1991}: stealth.ctf.
  \url{https://github.com/vexs1991/stealth-ctf/tree/export/}, accessed:
  2024-12-27

\bibitem{wazuh}
{Wazuh}: Open source siem. \url{https://wazuh.com/}, accessed: 2024-12-27

\bibitem{zafar24}
Zafar, A., Yamin, M., Katt, B., Torseth, E.: All flags are not created equal: A
  deep look into ctf scoring algorithms. Expert Systems with App.  (2024)

\bibitem{pwn2own}
{ZDI}: About. \url{https://www.zerodayinitiative.com/about/}, accessed:
  2024-12-27

\end{thebibliography}
\end{document}